\documentclass[%
 reprint, 
 aip, jcp,
nobibnotes,
]{revtex4-1}

\usepackage{amsmath}
\usepackage{xcolor}
\colorlet{mylinkcolor}{blue!66!black!80}
\usepackage[colorlinks=true,linkcolor=mylinkcolor,citecolor=mylinkcolor,filecolor=cyan,urlcolor=mylinkcolor,breaklinks=true]{hyperref}
\usepackage{mathtools}
\usepackage{physics}

\newcommand{\avg}[1]{\langle#1\rangle}
\newcommand{\bavg}[1]{\big\langle#1\big\rangle}

\newcommand{\e}{{\rm e}}

\begin{document}

\setcounter{page}{1} 

\title{Phase transition in thermodynamically consistent biochemical oscillators}
\author{Basile Nguyen}
\affiliation{II. Institut f{\"u}r Theoretische Physik, Universit{\"a}t Stuttgart, 70550 Stuttgart, Germany}
\author{Udo Seifert}
\affiliation{II. Institut f{\"u}r Theoretische Physik, Universit{\"a}t Stuttgart, 70550 Stuttgart, Germany}
\author{Andre C. Barato}
\affiliation{Max Planck Institute for the Physics of Complex Systems, N\"othnizer Strasse 38, 01187 Dresden, Germany}

\date{\today}%

\begin{abstract}
{
Biochemical oscillations are ubiquitous in living organisms. In an autonomous system, not influenced by an external signal, they can only occur out of equilibrium. We show that they emerge through a generic nonequilibrium phase transition, with a characteristic qualitative behavior at criticality. The control parameter is the thermodynamic force, which must be above a certain threshold for the onset of biochemical oscillations. This critical behavior is characterized by the thermodynamic flux associated with the thermodynamic force, its diffusion coefficient, and the stationary distribution of the oscillating chemical species. We discuss metrics for the precision of biochemical oscillations by comparing two observables, the Fano factor associated with the thermodynamic flux and the number of coherent oscillations. Since the Fano factor can be small even when there are no biochemical oscillations, we argue that the number of coherent oscillations is more appropriate to quantify the precision of biochemical oscillations. Our results are obtained with three thermodynamically consistent versions of known models: the Brusselator, the activator-inhibitor model, and a model for KaiC oscillations.
}
{}{}
\end{abstract}

\maketitle 

\section{INTRODUCTION}

Living systems need biochemical oscillators \cite{nov08} for the timing and control 
of several key processes such as circadian rhythms \cite{gold96,naka05} and 
the cell cycle \cite{ferr11}. Oscillations can only set in if the system is out of equilibrium. 
The control parameter that has to be non-zero for the system to be out of equilibrium is the thermodynamic force. 
For instance, for a system in contact with a bath that contains a fixed concentration 
of adenosine triphosphate (ATP), the thermodynamic force is the free energy liberated 
with the hydrolysis of one ATP molecule. 

More specifically, in a recent study on the relation between energy dissipation 
and the precision of biochemical oscillations,  Cao {\sl et al.}  \cite{cao15} 
have shown that the thermodynamic force must be above a certain threshold
for the onset of biochemical oscillations. Therefore, a natural question that 
arises is whether biochemical oscillators display a phase transition. 
In other words, what kind of non-analytical behavior do physical observables 
display at this critical thermodynamic force? It is worth noting that the relation 
between biochemical oscillations and thermodynamics has also been studied in 
\cite{gasp04,xiao08,vell09,rao11,bian14}. 

In this paper, we show that a generic phase transition takes place in biochemical 
oscillators. As in the well-known theory of 
nonequilibrium phase transitions \cite{nico78,schr79,bara82,walg83,nico86}, this phase transition is associated with a Hopf bifurcation, i.e., the onset of limit cycle, in the
deterministic rate equations. An observable that characterizes the transition is the steady state distribution of the chemical species that oscillates. This distribution becomes bimodal above the critical force. We analyze the critical behavior of the fluctuating thermodynamic current conjugate to 
the thermodynamic force. The average of this current is the rate at which the biochemical oscillator consumes ATP.   
The first derivative of this average with respect to the thermodynamic force is found to be discontinuous at 
the critical point. We also investigate fluctuations of this thermodynamic current. In particular, its diffusion 
coefficient is found to diverge there.  

Since these biochemical oscillations can occur in systems with a finite number of molecules that lead to relatively 
large fluctuations, it is natural to study the precision of biochemical oscillations \cite{cao15,bara17,wie18}.
More broadly, the relation between precision and dissipation in biophysics has been intensively investigated 
\cite{qian07,lan12,meht12,gove14a,hart15,ould17}. We analyze the relation between the Fano factor associated with 
the thermodynamic current and the number of coherent oscillations. The Fano factor has been analyzed for theoretical studies 
related to single molecule experiments \cite{schn95,chem08,moff14,bara15,bara15c} and has been proposed as an observable that can 
quantify the precision of biochemical oscillators \cite{bara16,wie18}. Interestingly, this Fano factor has a universal lower bound 
that depends solely on the thermodynamic force, which follows from the thermodynamic uncertainty relation \cite{bara15a,piet16,ging16}. 

The number of coherent oscillations, which is the number of periods for which different stochastic realizations remain coherent with each other, is a standard measure of the precision of biochemical oscillators \cite{cao15,bara17,qian00,gasp02,hou06,xiao07,more07,jorg18}. It can be used to identify the onset of biochemical oscillations, i.e., it is zero below the critical force and becomes non-zero above it \cite{cao15}. We show that the Fano factor does not properly quantify the precision of biochemical oscillations. This observable that diverges at the critical point, due to the divergence of the diffusion coefficient, is shown to be small below the critical point, indicating high precision even if there are no biochemical oscillations. 

We consider three different models for biochemical oscillators: the Brusselator \cite{nico77}, the activator-inhibitor model \cite{cao15}, and a model for the oscillations in the phosphorylation level of KaiC \cite{zon07}, which is a protein related to the regulation of the circadian rhythm of cyanobacteria \cite{dong08}. 

The paper is organized as follows. In Sec. \ref{sec:Crit} we introduce the three models and analyze their critical behavior. In Sec. \ref{sec:Precision} we discuss metrics for the precision of biochemical oscillations, with the comparison between the number of coherent oscillations and the Fano factor. We conclude in Sec. \ref{sec:Concl}. 
Details of the activator-inhibitor model and the KaiC model are provided in Appendix \ref{app_AI} and Appendix \ref{app_KaiC}, respectively.

\section{Phase transition} \label{sec:Crit}
\subsection{BRUSSELATOR} \label{sec:bruss}

\subsubsection{Model definition}

The Brusselator is a paradigmatic model  for biochemical oscillations \cite{nico77,lefe88,qian02,andr08}. It consists of two intermediate species 
$X$ and $Y$ in a volume $\Omega$. The external bath contains two chemical species $A$ and $B$ at fixed concentrations $[A]$ and $[B]$, respectively. The set of chemical reactions is
\begin{equation}
\begin{split}
  A &\xrightleftharpoons[{k_{-1}}]{k_{1}} X, \\
  B &\xrightleftharpoons[{k_{-2}}]{k_2} Y, \\
  2X+Y &\xrightleftharpoons[{k_3}]{k_3} 3X,
\end{split}
\label{eq:bruss_CME}
\end{equation}
where $k_1, k_{-1}, k_{2}, k_{-2}, k_3$ are transition rates. For convenience we assume that the transition
rates for the forward and backward direction in the third reaction are the same. The system is driven out of equilibrium 
due to a difference of chemical potential between $A$ and $B$,
which is written as  $\varDelta\mu \equiv \mu_B-\mu_A$. For example, consider the following cycle:  a $Y$ molecule is created with rate $k_2$, then a $Y$ molecule is 
transformed into an $X$ molecule with rate $k_3$ and, finally, an $X$ molecule is degraded with rate $k_{-1}$. This cycle leads to the consumption of substrate 
$B$ and generation of product $A$. The thermodynamic force associated with this cycle is  
 \begin{equation} 
 {\varDelta\mu} \equiv \ln \frac{k_{-1}k_2[B]}{k_{-2} k_1[A]},
 \label{eq:bruss_ltb}
\end{equation} 
where the temperature $T$ and Boltzmann's constant $k_B$ are set to $k_BT=1$ throughout this paper. 
The above relation between the thermodynamic parameter $\varDelta\mu$ and the transition rates is known as generalized detailed balance \cite{seif12}.

The state of the system is determined by two variables, the total number of $X$ molecules $n_X$ and the total number of $Y$ molecules $n_Y$. The time evolution of $P(n_{X},n_{Y},t)$, the probability 
 to find the system in state $(n_X, n_Y)$ at time $t$, is governed by the chemical master equation, which reads 

\begin{equation}
\begin{aligned}
&\partial_t P(n_{X},n_{Y},t) =\Big\{\Omega k_1 [A]\left[\epsilon_X^- - 1\right]  + \Omega k_2 [B]\left[\epsilon_Y^- - 1\right]  \\
& + k_{-1}\left[(n_X+1)\epsilon_X^+ - n_X\right]  + k_{-2}\left[(n_Y+1)\epsilon_Y^+ - n_Y\right] \\
& + \frac{k_3}{\Omega^2} \left[(n_X-1)(n_X-2)(n_Y+1)\epsilon_X^-\epsilon_Y^+ \right. \\
& - n_X(n_X-1)n_Y + (n_X+1)n_X(n_X-1)\epsilon_X^+\epsilon_Y^-  \\
& - \left. n_X(n_X-1)(n_X-2)\right] \Big\} P(n_{X},n_{Y},t),
\end{aligned}
\label{eq:bruss_CME2}
\end{equation}
where we define step operators as 
\begin{equation}
\begin{aligned}
\epsilon_{X}^\pm P(n_{X},n_{Y},t)&\equiv P(n_{X}\pm1,n_{Y},t), \\
 \epsilon_{Y}^\pm P(n_{X},n_{Y},t)&\equiv P(n_{X},n_{Y}\pm1,t). \\
 \end{aligned}
\end{equation}
The system reaches a 
nonequilibrium steady state with a steady-state distribution written as  $P(n_X,n_Y)$. The marginal distribution of $n_X$ that we evaluate in numerical simulations is defined 
as $P(n_X)\equiv \sum_{n_Y}P(n_X,n_Y)$

From the master equation \eqref{eq:bruss_CME2}, we obtain the equations for the time evolution of the densities 
\begin{equation}
\begin{aligned}
 x&\equiv\sum_{n_{X},n_{Y}}n_XP(n_{X},n_{Y},t)/\Omega, \\
  y&\equiv\sum_{n_{X},n_{Y}}n_YP(n_{X},n_{Y},t)/\Omega, \\
 \end{aligned}
 \end{equation}
 in the deterministic limit ($\Omega\rightarrow \infty$), which read
\begin{equation}
\begin{aligned}
\frac{d x}{dt} &= k_1[A] - k_{-1}x + k_3\left(x^2y - x^3\right),\\
\frac{d y}{dt} &= k_2[B] - k_{-2}y - k_3\left(x^2y - x^3\right).\\
\end{aligned}
\label{eq:bruss_Det}
\end{equation}

\begin{figure}
\centering
\includegraphics[width=.85\linewidth]{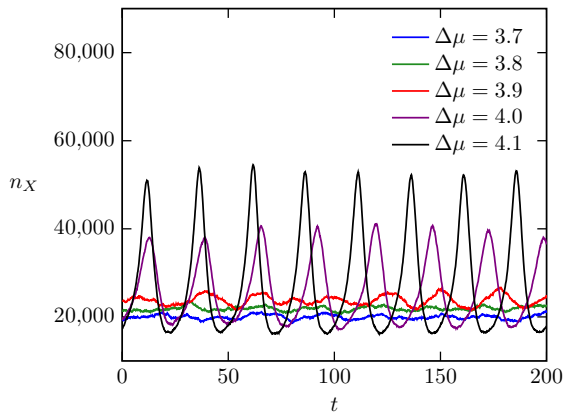}
\caption{Trajectories of $n_X$ at different $\varDelta\mu$ in the Brusselator with volume $\Omega=10^{5}$, the remaining parameters are given in the main text.}
\label{fig:bruss_CME_traj}
\end{figure}

We have performed continuous time Monte Carlo simulations using the Gillespie algorithm \cite{gill77}. 
We set the parameters to $k_1 = k_2 = 0.1$, $k_{-1}= k_3 = 1 $, $[A] = 1$, $[B] = 3$. The rate 
$k_{-2}$ is computed with $\varDelta\mu$ and the generalized detailed balance relation in Eq.~\eqref{eq:bruss_ltb},
where $\varDelta\mu$ is the control parameter. In Fig.~\ref{fig:bruss_CME_traj}, we show stochastic trajectories of this model. For large enough $\varDelta\mu$, biochemical oscillations set in.

\begin{figure}
\centering
\includegraphics[width=1\linewidth]{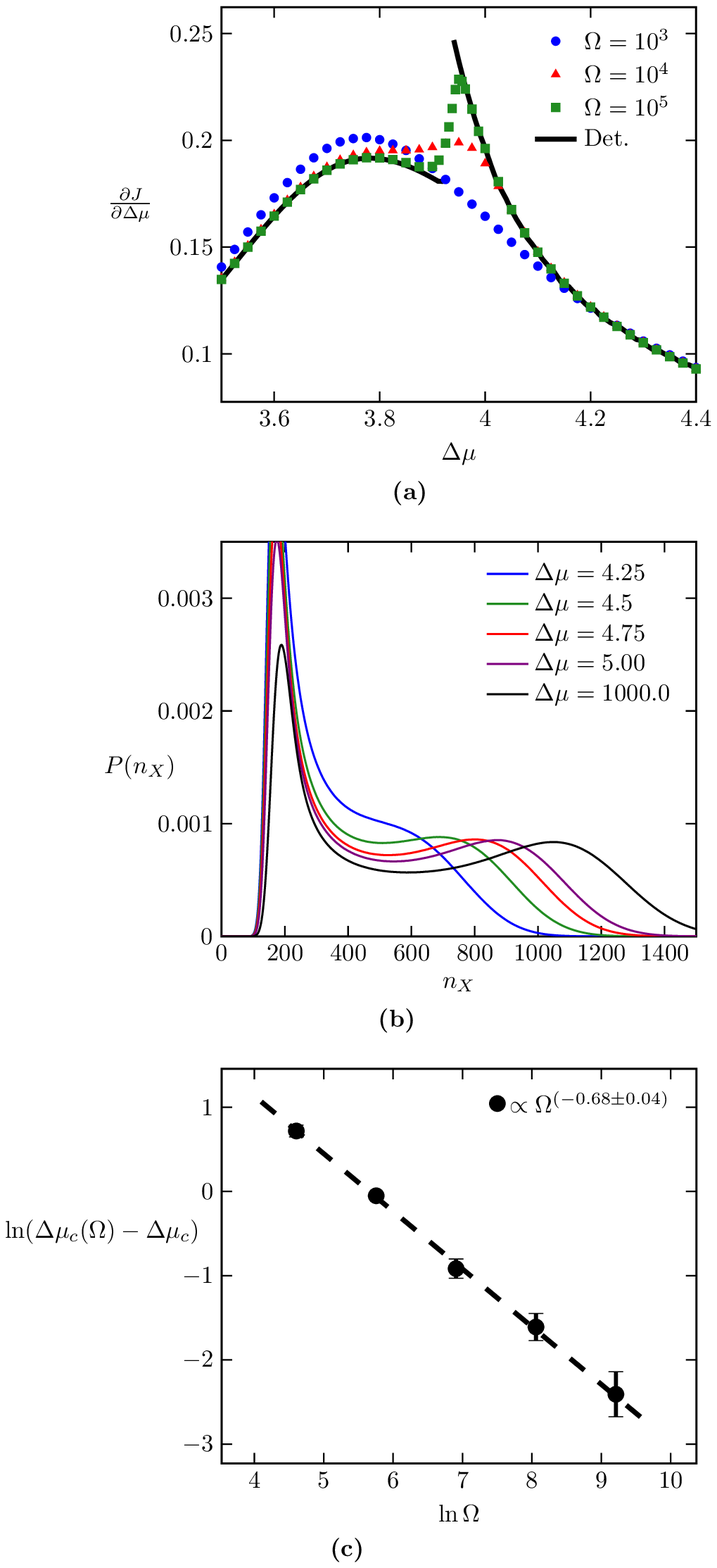}
\caption{Phase transition in the Brusselator. (a) First derivative of the thermodynamic flux $J$ as a function of  $\varDelta\mu$. The critical point is $\varDelta\mu_c\simeq 3.95$.  (b) Stationary distribution of chemical species $X$ for $\Omega = 10^{3}$ and different values of $\varDelta\mu$. (c) Difference between the point at which the distribution in (b) becomes bimodal in a finite system $\varDelta\mu_c(\Omega)$ and the critical point $\varDelta\mu_c\simeq 3.95$ obtained with the deterministic rate equations.}
\label{fig:bruss_CME_PJ}
\end{figure}

\begin{figure}
\centering
\includegraphics[width=1\linewidth]{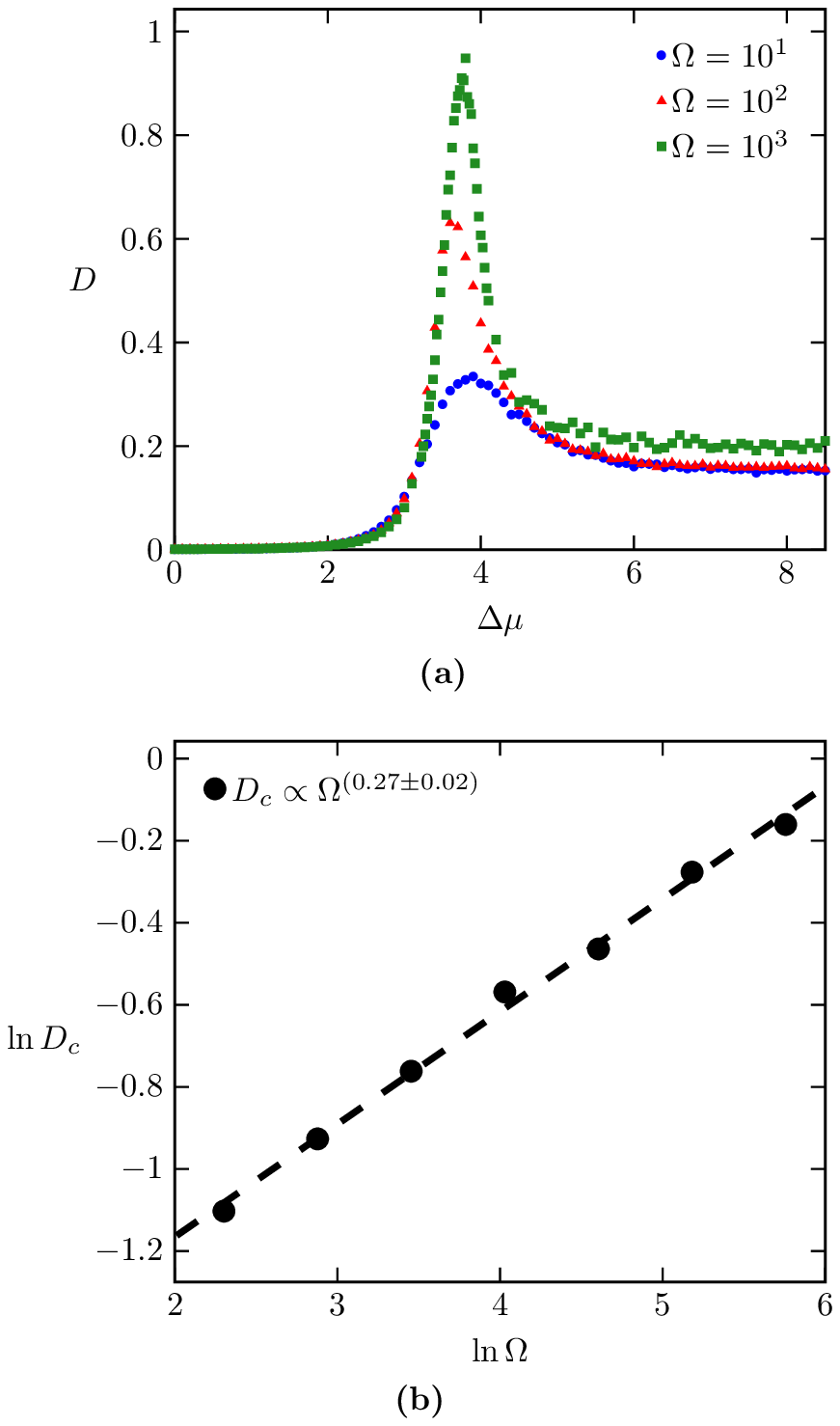}
\caption{Diffusion coefficient for the Brusselator. (a) Diffusion coefficient $D$ as a function of $\varDelta\mu$. 
(b) Maximum diffusion coefficient $D_c$ as a function of $\Omega$.}
\label{fig:bruss_CME_D}
\end{figure}


\subsubsection{Results}

In the deterministic limit described by Eq. \eqref{eq:bruss_Det}, for $\varDelta\mu\ge \varDelta\mu_c\simeq 3.95$ 
the stationary solution becomes numerically unstable and a limit cycle sets in. The thermodynamic flux associated with this force is the rate of generation of the product $A$ per volume $\Omega$, which must be equal the rate of consumption of $B$ due to the conservation of the number of particles in the reservoir. In the deterministic limit this flux takes the form 
\begin{equation}
\label{eq:det_J}
 J = k_{-1}x-k_1[A],
\end{equation}
where $x$ is the stationary solution of Eq. \eqref{eq:bruss_Det}. The rate of entropy production is simply given by $\sigma \equiv J \varDelta\mu$ in the steady state.  As shown in Fig.~\ref{fig:bruss_CME_PJ}(a), there is a discontinuity in the first derivative $\partial J / \partial \varDelta\mu$ at the critical point $\varDelta\mu_c$ in this deterministic limit.  

For a stochastic system, the average thermodynamic flux reads 
\begin{equation}
\label{eq:sto_J}
J= \Omega^{-1}k_{-1}\sum_{n_X}n_XP(n_X)-k_1[A].
\end{equation}
In Fig. \ref{fig:bruss_CME_PJ}(a), we plot $\partial J/\partial\varDelta\mu$ as a function of $\varDelta\mu$. With increasing system size,  the curve tends to the deterministic result, 
with an increase of $\partial J/\partial\varDelta\mu$ close to the critical point that gets steeper with increasing volume $\Omega$. 

For a finite system, there is a crossover to oscillatory behaviors, as illustrated in Fig.~\ref{fig:bruss_CME_traj}. The stationary probability distribution $P(n_X)$ also changes at this crossover. Below some finite-size critical point, where biochemical 
oscillations do not take place, this distribution is unimodal, whereas above this critical point this distribution becomes bimodal. This 
result is shown in Fig. \ref{fig:bruss_CME_PJ}(b). We define the finite-size critical point $\varDelta\mu_c(\Omega)$ as the minimal $\varDelta\mu$  for which the distribution displays a local minimum. In Fig. \ref{fig:bruss_CME_PJ}(c), we show that $\varDelta\mu_c(\Omega)$ converges to $\varDelta\mu_c\simeq 3.95$, where the difference $\varDelta\mu_c(\Omega)-\varDelta\mu_c$ decreases as a power-law with the system size $\Omega$. 

Fluctuations related to the thermodynamic flux can be analyzed by considering a stochastic time-integrated current $Z$, which is extensive in time. In a stochastic trajectory,  this random variable increases by one if an $A$ is produced, which happens if the transition with rate $k_{-1}$ takes place, and it decreases 
by one if an $A$ is consumed, which happens if the transition with rate $k_{1}$ takes place. The average flux in \eqref{eq:sto_J} can be defined as
\begin{equation}\label{eq:det_J2}
J\equiv \langle Z\rangle/(T\Omega),
\end{equation}
where $T$ is the time interval and the brackets denote an average over stochastic trajectories. 
This time interval is large enough compared to relaxation times so that the 
stationary regime is probed. The diffusion coefficient (per volume) associated with $Z$ is defined as
\begin{equation} \label{eq:bruss_D}
D \equiv \frac{\avg{Z^2} - \avg{Z}^2}{2T\Omega}.
\end{equation}  
In Fig.~\ref{fig:bruss_CME_D}(a) we show the diffusion coefficient $D$ as a function of $\varDelta\mu$. It has a maximum 
close to the critical point that increases with the volume $\Omega$. The maximum $D_c$ as a function of the 
volume $\Omega$ follows a power law with effective exponent $0.27\pm 0.02$, as shown in Fig.~\ref{fig:bruss_CME_D}(b). 
This finite-size scaling indicates that $D$ diverges at the critical point.


\subsection{ACTIVATOR-INHIBITOR MODEL} \label{sec:AI}
\label{sec3}

\begin{figure}
\centering
\includegraphics[width=1.0\linewidth]{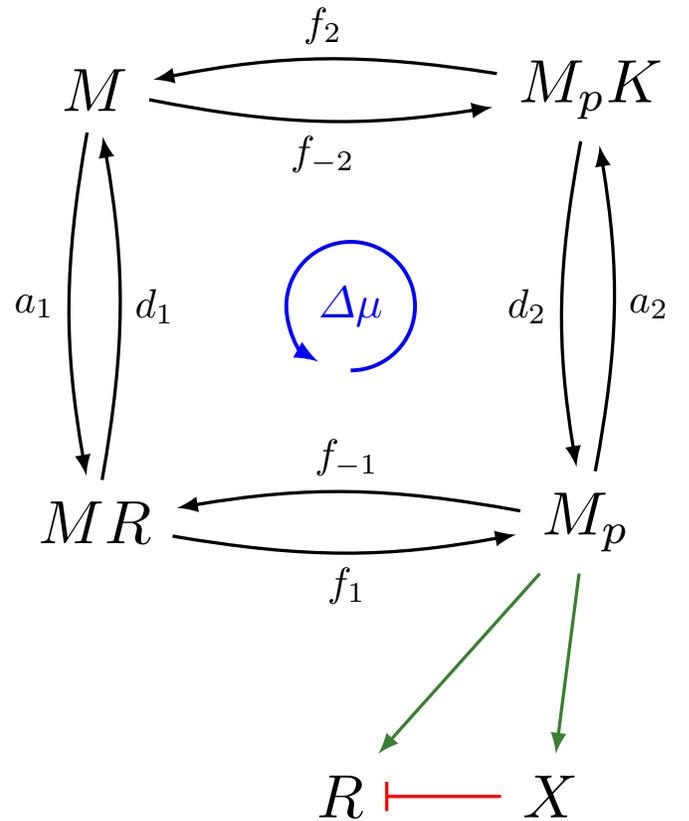}
\caption{Activator-inhibitor model. The transition rates for the phosphorylation cycle of an enzyme $M$ are represented with black arrows. 
The completion of the counter clock-wise cycle leads to the hydrolysis of one ATP, which liberates a free energy $\varDelta\mu$.
Green arrows represent activation and the red lines represent inhibition.
}
\label{fig:AI_model}
\end{figure}

\subsubsection{Model definition}

The activator-inhibitor model \cite{cao15} is a more elaborate biochemical oscillator compared to the Brusselator. The model is depicted in Fig.~\ref{fig:AI_model}. 
It consists of activators $R$, inhibitors $X$, enzymes $M$ and 
phosphatases $K$ interacting in a volume $\Omega$. The external bath contains fixed concentrations of ATP, ADP, and $\text{P}_\text{i}$. 
The enzyme $M$ can be in four different states ($M, MR, M_pK$, $M_p$),  where $M_p$ is the phosphorylated form of the enzyme.
For a phosphorylation reaction to take place, a $R$ molecule must be bound to the enzyme $M$ and for a dephosphorylation reaction to take place, a phosphatase $K$ must be bound to the enzyme $M$. In the phosphorylation reaction, one ATP is transformed into an ADP, whereas in the dephosphorylation reaction a $\text{P}_\text{i}$ is released in the solution. In the anti-clockwise cycle shown in Fig.~\ref{fig:AI_model}, one ATP is transformed into $\text{ADP}+\text{P}_\text{i}$, hence 
the thermodynamic force associated with this cycle is  
\begin{equation} \label{eq:AI_ltb}
\varDelta\mu=\mu_\text{ATP} - \mu_\text{ADP} - \mu_{\text{P}_\text{i}}\equiv\ln\frac{a_1 a_2 f_1 f_2}{d_1 d_2 f_{-1} f_{-2}},
\end{equation}
where the rates $a_i, d_i, f_{\pm i}$ are given in Fig.~\ref{fig:AI_model}.

Activators $R$ and inhibitors $X$ are related to this phosphorylation cycle 
in a feedback loop. The system contains a fixed concentration of substrate 
$S$, which is consumed (produced) when a $R$ or $X$ molecule is produced 
(consumed). The phosphorylated form of the enzyme $M_p$ catalyzes both the 
production of $R$ with a rate $k_0$ and the production of $X$ with a rate 
$k_3$ 
(positive feedback). Inhibitors $X$ degrade $R$ with a rate $k_2$ (negative feedback) and can be spontaneously degraded with a rate $k_4$. For thermodynamic consistency, we must include reverse rates, which are given by $\delta_i$, where $i=0,...,4$. These reactions correspond to the lower part of Fig.~\ref{fig:AI_model} and can be written as
\begin{equation}
\begin{aligned}
M_p+S&\xrightleftharpoons[\delta_0]{k_0}M_p+R, \\
S&\xrightleftharpoons[\delta_1]{k_1} R, \\
X+R+\textrm{ATP}&\xrightleftharpoons[\delta_2]{k_2} X+S+\textrm{ADP}+\textrm{P}_\textrm{i}, \\
M_p+S&\xrightleftharpoons[\delta_3]{k_3} M_p+X,\\
X+\textrm{ATP}&\xrightleftharpoons[\delta_4]{k_4}S+\textrm{ADP}+\textrm{P}_\textrm{i}, \\
\end{aligned}
\label{eq:AI_eq} 
\end{equation}
where generalized detailed balance \cite{seif12} requires
\begin{equation}
\begin{aligned}
\delta_1 &= k_1 \delta_0/k_0,\\
\delta_2 &= \e^{-\varDelta\mu} k_2 k_0/\delta_0, \\
\delta_4 &= \e^{-\varDelta\mu} k_4 k_3/\delta_3. \\
\end{aligned}
\label{eq:AI_ltb} 
\end{equation}
In contrast to the model from \cite{cao15}, we have added ATP consumption 
in the chemical reactions Eq.~\eqref{eq:AI_eq} for thermodynamic 
consistency. 

An additional feature of the activator-inhibitor model in relation to the Brusselator is the competition for a scarce number of phosphatases $n_{K_{\textrm{tot}}}$. In order for oscillations to set in, 
this number must be at some intermediate optimal value. If $n_{K_{\textrm{tot}}}> n_{M_{\textrm{tot}}}$, then the phosphorylation cycle shown in Fig.~\ref{fig:AI_model} for different 
enzymes does not synchronize, since there is always free phosphatase to bind to the enzyme, which is necessary for the dephosphorylation reaction. If $n_{K_{\textrm{tot}}}$
is too small then only a few enzymes can complete their cycle in a synchronized way. This competition for a scarce resource is a common feature in more realistic biochemical oscillators, such as the model for KaiC oscillations in the next section.

The chemical master equation and the respective deterministic rate equations for this model are shown in Appendix $\ref{app_AI}$. The state of the system is determined by a vector of the numbers of molecules $n_i$, with $i=R, X , K , M , MR, M_p, M_pK$. 
This vector is subjected to the constraints $n_{M_{\textrm{tot}}}=n_{M}+ n_{MR}+ n_{M_p}+ n_{M_pK}$ and $n_{K_{\textrm{tot}}}=n_{K}+n_{M_pK}$, where $n_{M_{tot}}$ is the total number of enzymes 
and $n_{K_{tot}}$ is the total number of phosphatases. The volume of the system is $\Omega$ and concentrations are denoted by $[i]\equiv n_i/\Omega$. The concentration  of 
enzymes is set to $[M_{tot}]= 10$, the concentration of phosphatases is $[K_{tot}]=0.8$ and the concentration of substrates is $[S]=1$. The rates are set to $k_0=k_2=k_3=0.02, k_1=0.008, k_4=0.01, \delta_0=\delta_3=0.001, f_1=f_2=d_1=d_2=0.3, a_1=a_2=2.$ The rates $\delta_1,\delta_2,\delta_4$ are computed with the generalized detailed balance relation in Eq.~\eqref{eq:AI_ltb} and $f_{-1} = f_{-2} =  2e^{-\varDelta\mu/2}$, where $\varDelta\mu$ is the control parameter. 

The chemical species that we observe in our numerical simulations is $X$, which, depending on $\varDelta\mu$, can display biochemical oscillations. 
The fluctuating thermodynamic time-integrated current $Z$ in this model is the total number of ATP consumed: if the reaction with rate $f_1$ $(f_{-1})$ in Fig.~\ref{fig:AI_model}
takes place, the $Z$ increases (decreases) by one. In addition, if the 
reactions with rate $k_2,k_4$ $(\delta_2,\delta_4)$ in Eq.~\eqref{eq:AI_eq} 
take place, $Z$ also increases (decreases) by one.

The average flux $J$ and diffusion coefficient per volume $D$ are defined as 
in Eq. \eqref{eq:det_J} and Eq. \eqref{eq:bruss_D}, respectively. 

\subsubsection{Results}


\begin{figure}
\centering
\includegraphics[width=1\linewidth]{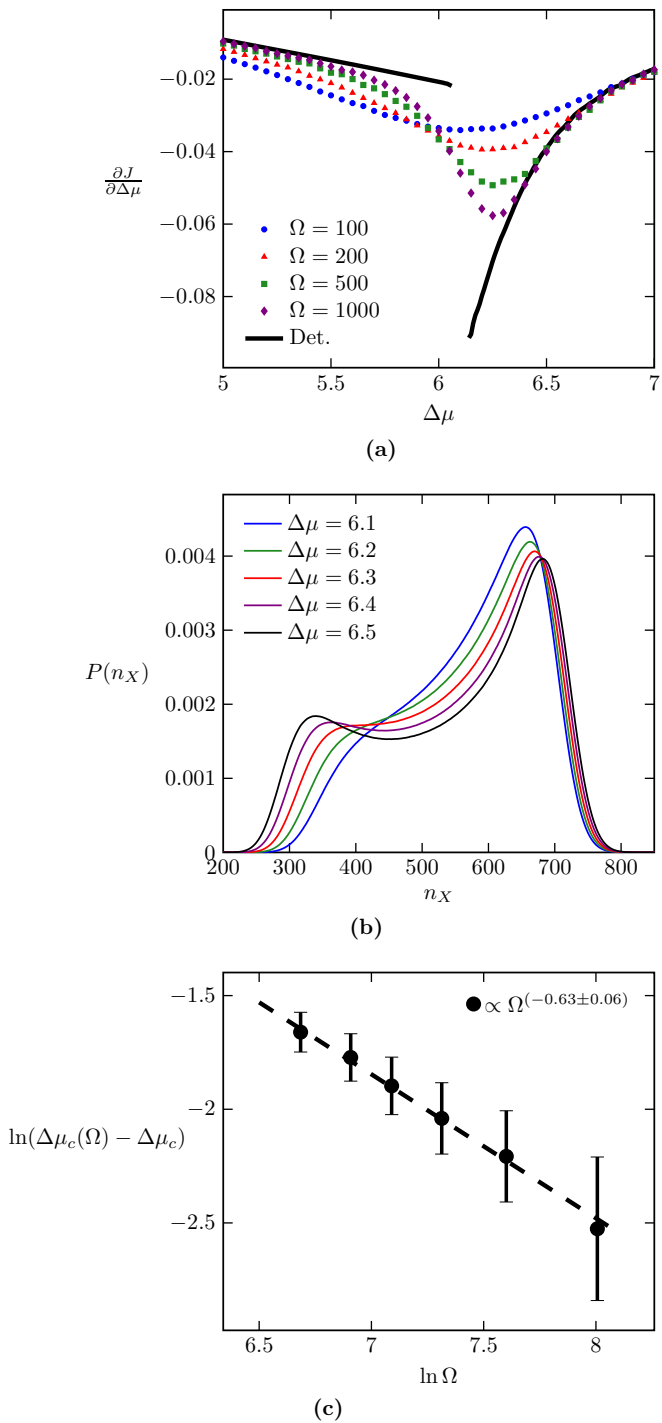}
\caption{Phase transition for the activator-inhibitor model. (a) First derivative of the flux $\partial J/\partial \varDelta\mu$ as a function of $\varDelta\mu$. The critical point is $\varDelta\mu_c\simeq 6.1$. 
(b) Stationary probability distribution of $n_X$ for different values of $\varDelta\mu$ and  $\Omega = 100$.  
The distribution becomes bimodal above the critical point, which is $\varDelta\mu \simeq 6.3$ for this finite $\Omega$. (c) Difference between the point at which the distribution in (b) becomes bimodal in a finite system $\varDelta\mu_c(\Omega)$ and the critical point $\varDelta\mu_c\simeq 6.1$ obtained with the deterministic rate equations.
}
\label{fig:AI_CME_PJ}
\end{figure}

\begin{figure}
\centering
\includegraphics[width=.83\linewidth]{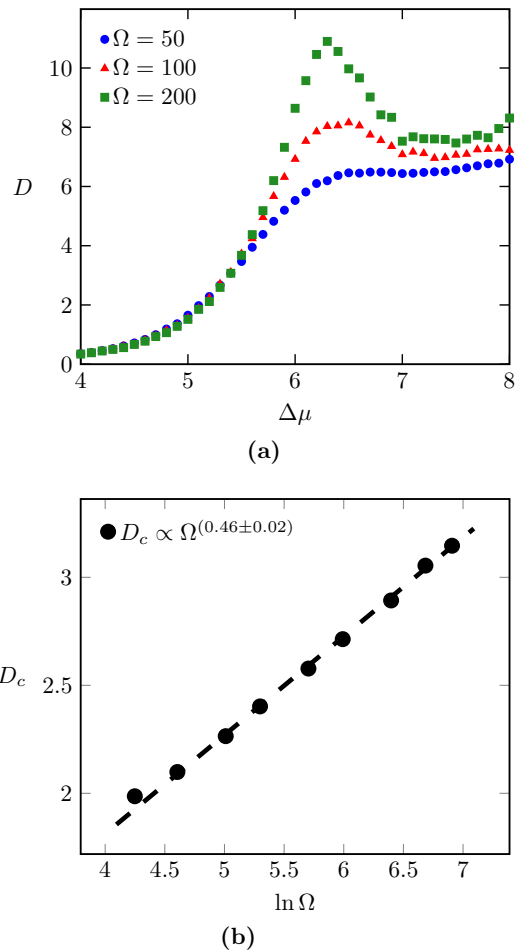}
\caption{Diffusion coefficient for the activator-inhibitor model. (a) Diffusion coefficient $D$ as a function of $\varDelta\mu$. (b) 
Maximum diffusion coefficient $D_c$ as a function of $\Omega$.}
\label{fig:AI_CME_D}
\end{figure}

As shown in Fig. \ref{fig:AI_CME_PJ}, the critical behavior of the activator-inhibitor model is qualitatively similar to the Brusselator. The first derivative of the flux $\partial J/\partial\varDelta\mu$ has a discontinuity at the critical point in the 
thermodynamic limit, as obtained from the deterministic rate equations shown in Appendix $\ref{app_AI}$. Furthermore, the stationary distribution $P(n_X)$ becomes bimodal above the critical point, which depends on the system size. The diffusion coefficient $D$ diverges at the critical point, as shown in Fig. \ref{fig:AI_CME_D}. The effective exponent related to the finite-size scaling of the maximum diffusion coefficient $D_c$ is ${0.46\pm 0.02}$, which is different from the one found in the Brusselator.

\subsection{KAIC MODEL} \label{sec:KaiC}

\begin{figure}
\centering
\includegraphics[width=1.0\linewidth]{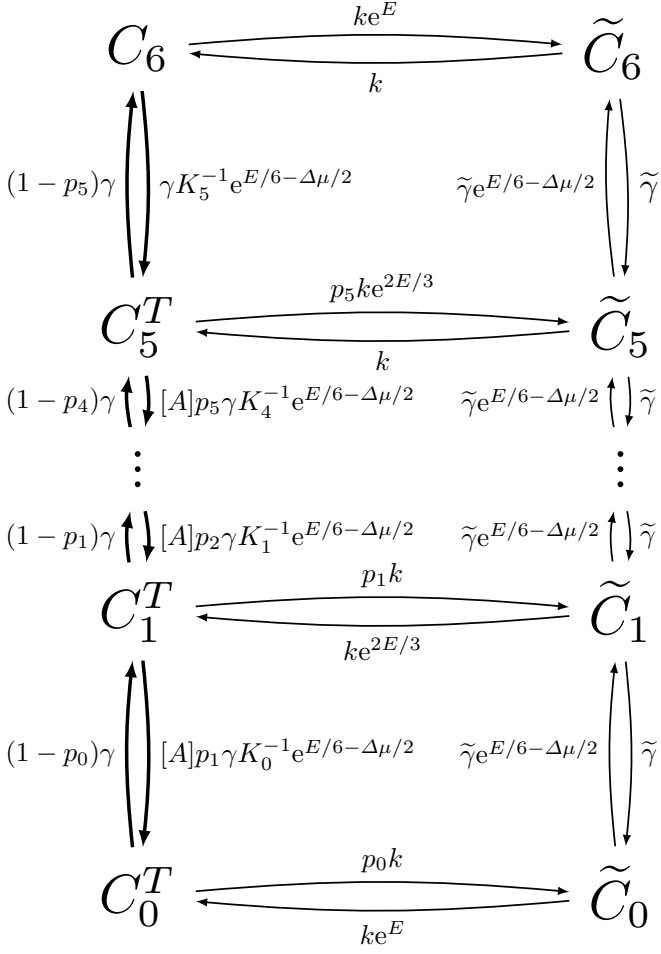}
\caption{KaiC model with differential affinity. $C_j^T = C_j + AC_j$ are the active KaiC states with $j$ bound phosphates, which can be bound to a KaiA($A$). We assume fast binding and unbinding of KaiA with $C_j$, which is characterized by a dissociation constant $K_j=K_0 a^j$ and the fraction of unbound states $p_j = K_j/(K_j+[A])$. Note that the fully phosphorylated state $C_6$ cannot bind to $A$. $\widetilde{C}_j$ are the inactive states with $j$ bound phosphates. The transition rate from $C_j$ to $\widetilde{C}_j$ is $k\e^{\chi E(j-3)/3}$ and the transition rate from $\widetilde{C}_j$ to $C_j$ is $k\e^{\widetilde{\chi}E(3-j)/3}$, where $\chi$ ($\tilde{\chi}$) is an indicator function that is 0 (1) for $j=0,1,2,3$ and 1(0) for $j=4,5,6$. See Appendix \ref{app_KaiC} for the chemical master equation and rate equations.}
\label{fig:KaiC_model}
\end{figure}

\begin{figure}
\centering
\includegraphics[width=1\linewidth]{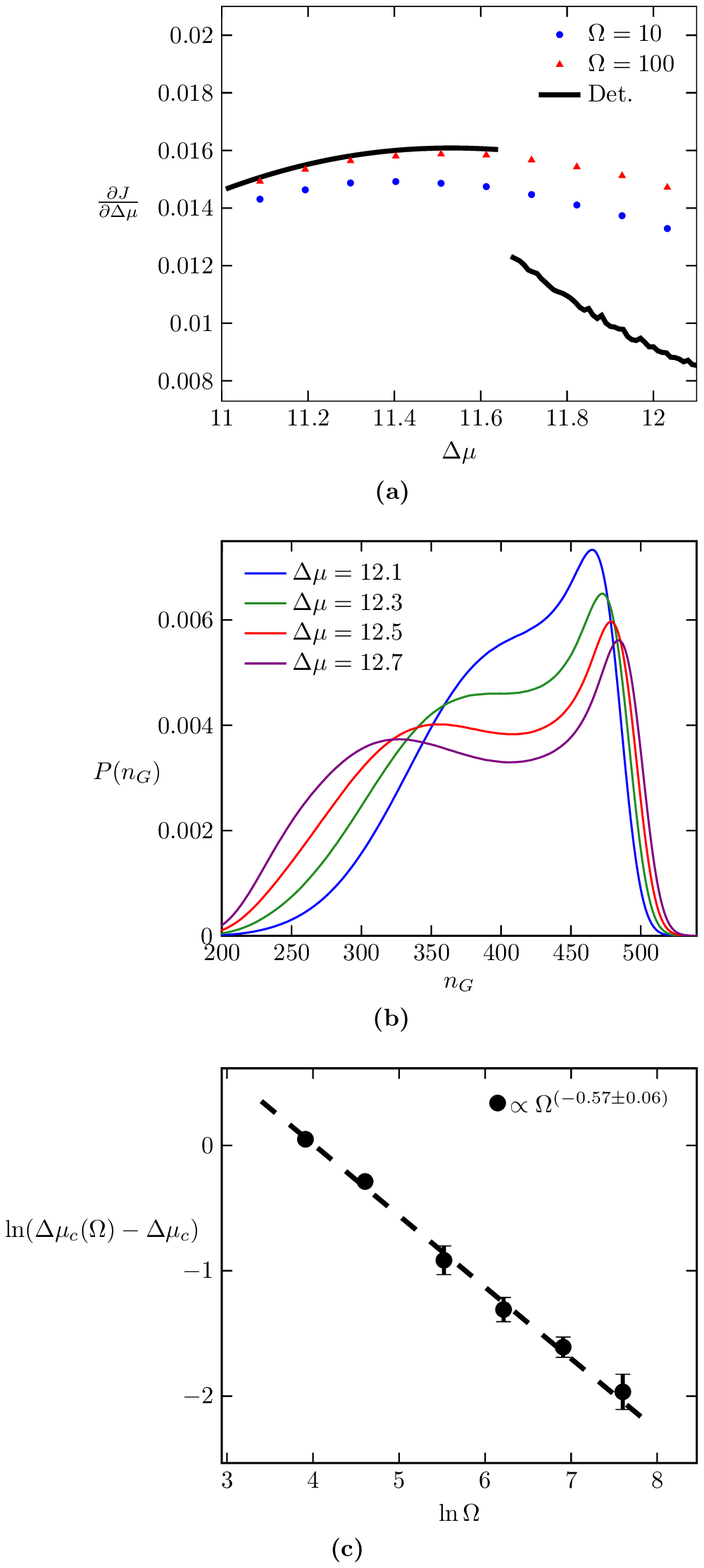}
\caption{Phase transition for the KaiC model. (a) First derivative of the flux $\partial J/\partial \varDelta\mu$ as a function of $\varDelta\mu$. The critical point is $\varDelta\mu_c \simeq 11.6$. 
(b) Stationary probability distribution of the phosphorylation level $n_G$ for different values of $\varDelta\mu$ and  $\Omega = 100$.  
The distribution becomes bimodal above the critical point, which is $\varDelta\mu \simeq 12.3$ for this finite $\Omega$. (c) Difference between the point at which the distribution in (b) becomes bimodal in a finite system $\varDelta\mu_c(\Omega)$ and the critical point $\varDelta\mu_c \simeq 11.6$ obtained with the deterministic rate equations. 
}
\label{fig:KaiC_CME_PJ}
\end{figure}

\begin{figure}
\centering
\includegraphics[width=1\linewidth]{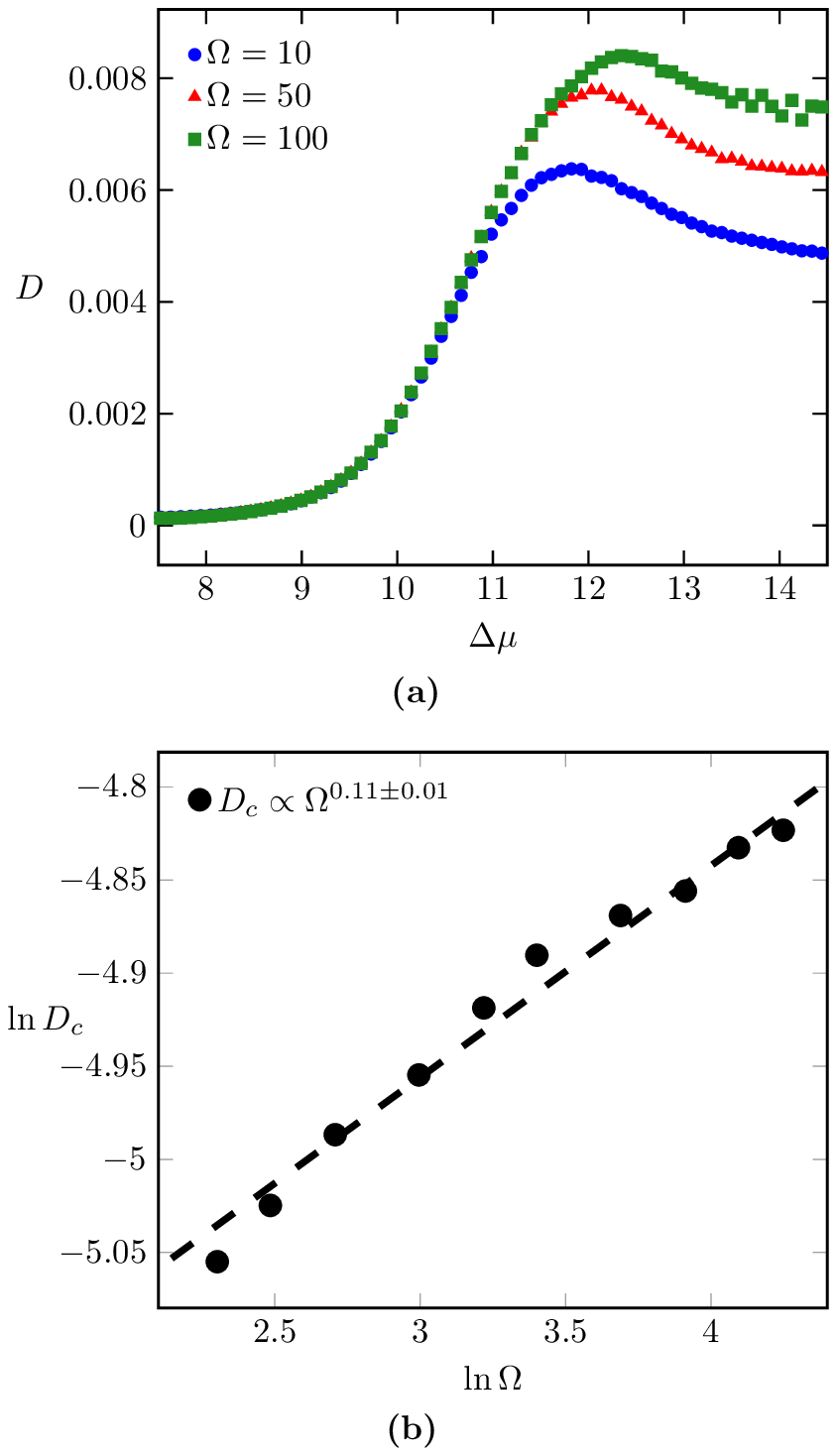}
\caption{Diffusion coefficient for the KaiC model. (a) Diffusion coefficient $D$ as a function of $\varDelta\mu$. (b) 
Maximum diffusion coefficient $D_c$ as a function of $\Omega$.}
\label{fig:KaiC_CME_D}
\end{figure}

\subsubsection{Model definition} 

There are several models for the oscillations of the phosphorylation level of KaiC proteins (see \cite{paij17} for a summary). 
Here, we analyze a modified version of a model introduced in \cite{zon07}. In particular, we make all transitions reversible for 
thermodynamic consistency.

The model contains KaiC molecules and KaiA molecules in a volume $\Omega$. Each KaiC molecule can be in 
14 different states denoted by $C_j$ and $\widetilde{C}_j$, where $j=0,1,\ldots,6$. The variables $j$ indicates the 
phosphorylation level of the molecule, which has 6 phosphorylation sites.  The state $C_j$ indicates the molecule is 
active and the state $\widetilde{C}_j$ indicates the molecule is inactive. The free energy of state $C_j$ minus the free 
energy of state $\widetilde{C}_j$ is given by 
\begin{equation}
\varDelta E_j=-E+jE/3. 
\end{equation}
If the molecule is active then a phosphorylation reaction can happen and if the molecule is inactive a dephosphorylation reaction can happen. 

An essential feature of the model is that a KaiA molecule must bind to the KaiC molecule for a phosphorylation reaction. The KaiA  
molecules play a role similar to the phosphatases in the activator-inhibitor model, i.e., it is a scarce resource that synchronizes
the phosphorylation cycle of different KaiC molecules. The dissociation constant for the binding of an A to a KaiC molecule in state $C_j$ is 
$K_j=K_0 a^j$. The constant $a>1$ makes the dissociation constant an increasing function of $j$, which is necessary for the 
onset of biochemical oscillations \cite{zon07}.

The transition rates for the model are as follows. KaiA(A) can bind with rate $\alpha$ to active states which are partially 
phosphorylated ($j=0,...,5$) to form a complex $AC_j$. The transition rate for the unbinding of $A$ from a KaiC in state $AC_j$ is 
$\alpha K_j$. The rate for a phosphorylation reaction from $AC_j$ to $C_{j+1}$ is $\gamma$, while the rate for the reverse reaction is rate $\gamma K_j^{-1} \e^{E/6-\varDelta\mu/2}$. 
Dephosphorylation from $\widetilde{C}_{j+1}$ to $\widetilde{C}_j$  occurs with rate $\widetilde{\gamma}$ , whereas the rate for the reversed reaction is $\widetilde{\gamma}\e^{E/6-\varDelta\mu/2}$. 
The transition rate from $C_j$ to $\widetilde{C}_j$ is $k\e^{\chi E(j-3)/3}$ and the transition rate from $\widetilde{C}_j$ to $C_j$ is $k\e^{(1-\chi)E(3-j)/3}$, where $\chi_j$ 
is an indicator function that is 0 for $j=0,1,2,3$ and 1 for $j=4,5,6$. This transition rates are thermodynamically consistent since the affinity of a phosporylation cycle is $\varDelta\mu$.  The chemical master equation for this model is shown in Appendix $\ref{app_KaiC}$.

We assume fast binding and unbinding kinetics of KaiA and introduce coarse-grained states $C_j^T = C_j + AC_j$ for $j=0,..,5$. The transition rates for this coarse-grained model are shown in Fig. \ref{fig:KaiC_model}. Within this assumption of 
timescale separation, the entropy production associated with this 
coarse-grained model remains the same as the entropy production of 
the full model \cite{espo12}. The fraction of unbound states among $C^T_j$ is $p_j = K_j/(K_j+[A])$,
where [A] is the concentration of free KaiA. This concentration fulfills the constraint  
\begin{equation}
\left[A_\text{tot}\right] = \left[A\right]+\sum_{j=0}^6 \frac{\left[A\right] \left[C_j^T\right]}{K_{j}+\left[A\right]},
\label{eq:KaiA_tot}
\end{equation}
where $\left[A_\text{tot}\right]\Omega$ is the total number of KaiA molecules. The total number of KaiC is $[C_\text{tot}]\Omega$, which leads to the constraint
\begin{equation}
[C_\text{tot}]= \sum_{j=0}^6 \left([C_j] + [\widetilde{C_j}]\right).
\label{eq:KaiC_tot}
\end{equation}
The chemical master equation and deterministic rate equations for this coarse-grained model are also shown in Appendix $\ref{app_KaiC}$.

The concentration of KaiC is set to $[C_\text{tot}] = 1$, while the concentration of KaiA  is set to $[A_\text{tot}]=0.04$. The parameters determining the transition rates are set to 
$\gamma=10, \widetilde{\gamma} = 1, E=20, k=e^{-E}, a=10, K_0=(1/3)10^{-7}$. The thermodynamic force $\varDelta \mu$ is kept as a free parameter. In our simulations, we  
monitor the phosphorylation level of the KaiC system, which is defined as  
\begin{equation} \label{eq:phosph}
n_G = \sum_{j=1}^6 j \left( [C_j] + [\widetilde{C}_j] \right).
\end{equation}
Depending on $\varDelta\mu$ this quantity can exhibit oscillations. The probability distribution $P(n_G)$ is the steady state probability of this observable. 
The fluctuating thermodynamic time-integrated current $Z$ in this model is the total of number of ATP consumed. For the KaiC molecule in the active state,
if a transition that increases $j$ takes place $Z$ increases by one and if the reversed transition takes place $Z$ decreases by one. 

\subsubsection{Results}

The critical behavior of the present model is  similar to the critical behavior observed with the Brusselator and the activator-inhibitor model, as shown in Fig.~\ref{fig:KaiC_CME_PJ}. 
The first derivative of the flux $\partial J/\partial \varDelta\mu$ has a discontinuity at the critical point in the thermodynamic limit and  
the stationary distribution $P(n_G)$ becomes bimodal above the critical point. For this model, concerning the quantity $\partial J/\partial \Delta\mu$, we could not numerically access systems sizes that are large enough for a better agreement between finite systems and the result from deterministic rate 
equations in Fig.~\ref{fig:KaiC_CME_PJ}(a). The diffusion coefficient $D$ diverges at the critical point, as shown in Fig.~\ref{fig:KaiC_CME_D}. 
The effective exponent related to the finite-size scaling of the maximum $D_c$ as a function of the the volume $\Omega$ is ${0.11\pm 0.01}$, which is smaller then the effective exponents 
found for the other two models. 

\section{Criteria for the precision of biochemical oscillations} \label{sec:Precision}
\begin{figure}
\centering
\includegraphics[width=1\linewidth]{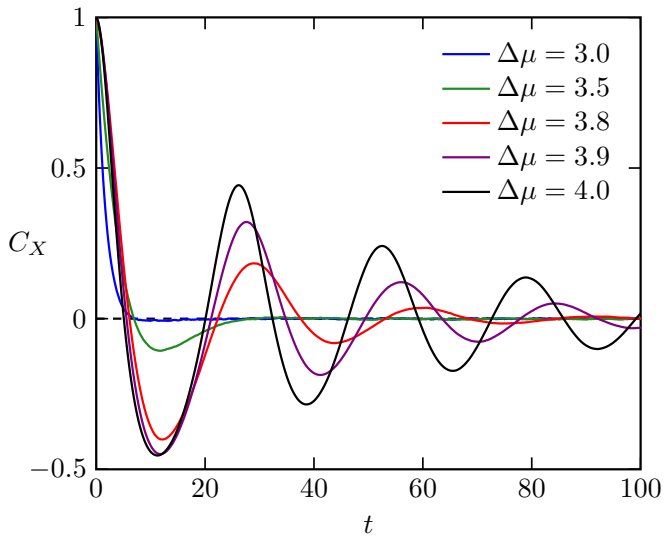}
\caption{Correlation function for species X (Eq. \eqref{eq:bruss_corr}) in the Brusselator for different $\varDelta\mu$.  
The volume is $\Omega=10^{3}$.}
\label{fig:bruss_CME_C}
\end{figure}

\begin{figure}
\centering
\includegraphics[width=1\linewidth]{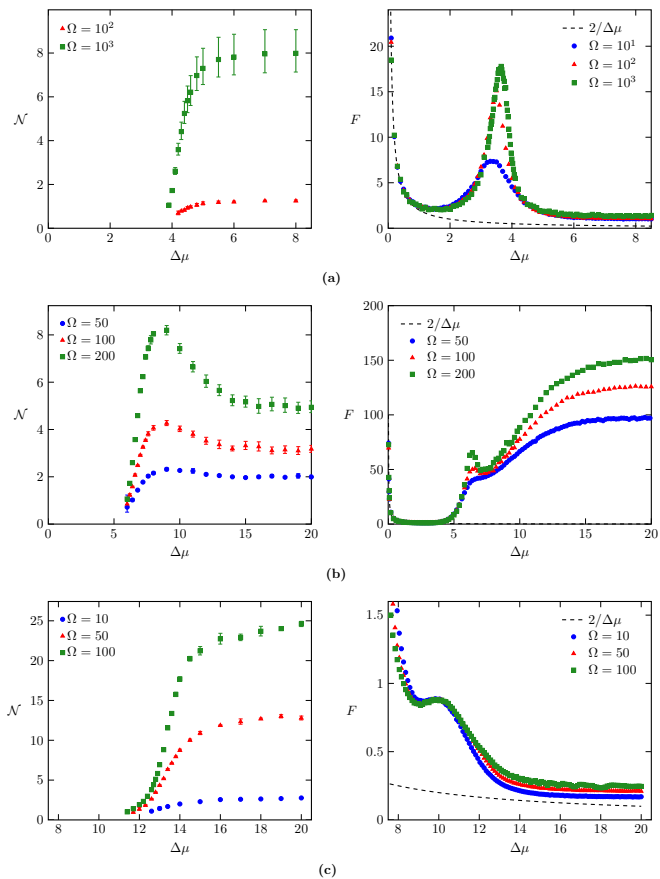}
\caption{Comparison between the number of coherent oscillations 
$\mathcal{N}$ and the Fano factor $F$. (a) Brusselator ($\mathcal{N}$ could 
not be measured for $\Omega=10$). (b) Activator-inhibitor model. (c) Model 
for KaiC oscillations. The right panels also show the lower bound 
\eqref{eq:bruss_R} for $F$}
\label{fig:CME_RF}
\end{figure}

Biochemical oscillations can occur in finite systems that display large fluctuations. In this section, we compare two quantities 
that can quantify the precision of biochemical oscillations, the number of coherent oscillations and the Fano factor associated with the thermodynamic flux.
The first quantity is related to the correlation function
\begin{equation} \label{eq:bruss_corr} 
 C_X(t) \equiv \bavg{\left( n_X(t) - \avg{n_X} \right) \left(n_X(0) - \avg{n_X} \right)}.
\end{equation}
where $n_X(t)$ is the number of $X$ molecules at time $t$, the average number of molecules is 
$\avg{n_X}\equiv \sum_{n_X}n_XP(n_X)$. Note that for the KaiC model the same expression is
valid with $n_G$ instead of $n_X$. In our simulations of all three models, the system starts
at the initial time $t=0$ in the stationary state. In Fig. \ref{fig:bruss_CME_C}, we show the correlation function for species $X$ in the Brusselator. Above the critical point, it displays oscillations with an exponentially decreasing amplitude . The number of coherent oscillations $\mathcal{N}$ is defined as the decay time divided by the period. It gives the typical number of 
oscillations for which two different stochastic realizations would remain coherent with each other. Therefore, a non-zero  $\mathcal{N}$
is a signature of biochemical oscillations, since it is a result of a non-zero imaginary part of the first excited eigenvalue of the stochastic matrix \cite{qian00,bara17}.

As shown in Fig.~\ref{fig:CME_RF}, for a finite system, $\mathcal{N}$ becomes non-zero above some critical value of $\varDelta\mu$. It is difficult to evaluate $\mathcal{N}$ numerically close to the critical point. If the number of coherent oscillations is too small,
it is not possible to obtain the period of oscillation from direct numerical evaluation of the correlation function.   
Hence, with our numerical results, we cannot determine whether $\mathcal{N}$ has a jump 
or approaches zero smoothly at the critical point. However, for simple three-state models it is possible to evaluate $\mathcal{N}$ 
by calculating the first non-zero eigenvalue of the stochastic matrix \cite{bara17,qian00}. In this analytical case, 
$\mathcal{N}$ approaches zero smoothly, without a discontinuity. Within our numerical simulations, the first $\varDelta\mu$ for which $\mathcal{N}$ becomes non-zero is smaller then the critical $\varDelta\mu$ for which the probability distribution of the oscillating species becomes bimodal, for all three models. 

For the Brusselator, the scaling of $\mathcal{N}$  with the volume $\Omega$ has been analyzed for both $\Delta\mu$ close to the critical point \cite{hou06,xiao07} and $\Delta\mu$ above the critical point \cite{gasp02}. In the first case, the scaling $\mathcal{N}\propto\Omega^{1/2}$ has been observed, while in the second $\mathcal{N}\propto\Omega^{1}$ has been obtained with the linear noise approximation. We have observed that close to the critical point our results for all three models, which are shown in Fig.~\ref{fig:CME_RF}, agree with the scaling $\mathcal{N}\propto\Omega^{1/2}$. Above the critical point, we find an exponent smaller than $1$, which suggest that the linear noise approximation, which is valid for large $\Omega$, does not apply to the system sizes we have considered. Furthermore, in \cite{xiao07} Xiao et al show that $\mathcal{N}$ becomes non-zero below the critical point of the Brusselator. We also observe a non-zero $\mathcal{N}$ below $\varDelta\mu_c$ for the activator-inhibitor and KaiC models.


The Fano factor $F$ associated with the fluctuating current $Z$ is defined as 
\begin{equation}    \label{eq:bruss_F}
 F \equiv \frac{\avg{Z^2} - \avg{Z}^2}{\avg{Z}}=\frac{2D}{J},
\end{equation}
which is a dimensionless parameter that quantifies the fluctuations of the 
thermodynamic time-integrated current $Z$.
A small Fano factor means that the current is precise. In equilibrium, where there are no 
biochemical oscillations,  the Fano factor diverges, since the average current is zero. Hence, this quantity is 
consistent with the absence of biochemical oscillations in equilibrium. Due to the divergence 
of the diffusion coefficient $D$, the Fano factor also diverges at the critical point in the thermodynamic limit.  

The thermodynamic uncertainty relation\cite{bara15a} implies the bound
\begin{equation} \label{eq:bruss_R}
F\ge \frac{2}{\varDelta \mu}.
\end{equation}
The highest precision achievable increases with the thermodynamic force $\varDelta \mu$. Therefore, a large $\varDelta \mu$ is 
a necessary condition for a small Fano factor. This inequality is illustrated in Fig.~\ref{fig:CME_RF}. Close to the 
critical point, $F$ is substantially larger than the bound with a local maximum that increases with the volume $\Omega$. 

Two main motivations to consider the Fano factor as a metric for the precision of biochemical oscillations are as follow. First, in light of the thermodynamic uncertainty relation, the Fano factor is a natural observable to relate precision with thermodynamics. Second, for a simple unicyclic mode, the Fano factor can be related to the cycle completion time (i.e. the period of oscillations) \cite{wie18}. Furthermore, the Fano factor associated with the ATP consumption seems particularly appealing to quantifying the precision of oscillations in the KaiC model, for which we consider oscillations of the phosphorylation level.  

However, our results in Fig.~\ref{fig:CME_RF} indicate that $F$ does not appropriately quantify the precision of biochemical oscillations. In particular, for the activator-inhibitor in Fig.~\ref{fig:CME_RF}(b), the Fano factor is minimal below the critical point where there are no biochemical oscillations. In general, since the Fano factor diverges at the critical point in the thermodynamic limit, a system that is large enough should display a smaller $F$ below the critical point as compared to 
the $F$ is some region above the critical point. Note that it is more difficult to observe this effect in the KaiC model, due to the smaller effective exponent associated with 
the divergence of $D$. For the Brusselator and KaiC model, above the crossover to oscillatory behavior, an increase in $\varDelta\mu$ leads to 
an increase in $\mathcal{N}$ and a decrease in $F$. However, for the activator-inhibitor model, there is a region around $\varDelta\mu \simeq 8$ 
where both $\mathcal{N}$ and $F$ increase with $\varDelta\mu$. Hence, for this last model, even if we restrict to a region where biochemical 
oscillations exist, $F$ does not appropriately correlate with the precision quantified by  $\mathcal{N}$.

\section{CONCLUSION} \label{sec:Concl}

We have characterized  a generic phase transition in biochemical oscillators. A control parameter for this phase transition is
the thermodynamic force that drives the system out of equilibrium. The stationary distribution of the oscillating chemical species 
becomes bimodal above the critical force, the first derivative of the thermodynamic flux with respect to the force is discontinuous 
at criticality, the diffusion coefficient (and Fano factor) associated with the thermodynamic flux diverges at criticality, and the 
number of coherent oscillations, which is a signature of biochemical oscillations, becomes nonzero above the critical point. We
have estimated effective exponents for the divergence of the diffusion coefficient with a limited range of volumes, a more 
reliable calculation of this exponent remains an open problem.

Three different models that display a limit cycle in the deterministic limit have been investigated. We expect that models for biochemical oscillators with this feature display qualitatively the same critical behavior. For models with purely stochastic biochemical oscillations \cite{McKa07}, i.e., models that do not have a limit cycle in the deterministic limit, this phase transition remains unexplored so far. The critical behavior of the entropy production has also been investigated in several different models with nonequilibrium phase transitions \cite{croc05,andr10,oliv11,tom12,bara12,zhan16,fala18}. The present case constitutes an example for which the first derivative of entropy production, which is the flux multiplied by the force, displays a discontinuity at the phase transition. We expect that the diffusion coefficient associated with the fluctuating entropy production diverges at the critical point for other models with nonequilibrium phase transitions. 

The precision of biochemical oscillations can be characterized by the Fano factor $F$ associated with the thermodynamic flux (or related first passage variables \cite{wie18}) and the number of coherent oscillations $\mathcal{N}$. We have shown that the Fano factor $F$ can indicate high precision even if there are no biochemical oscillations. This result suggests that $F$ does not quantify the precision of biochemical oscillations in a reliable way. We anticipate that this problem will also happen with probability currents that are different from the thermodynamic flux, which could be considered as more suitable for quantifying the precision of biochemical oscillations, since their diffusion coefficient will also diverge at criticality.

It remains to be seen whether and how the phase transition that we have characterized here can be used to understand 
the operation and design principles of biochemical oscillators. We have restricted our study to autonomous biochemical 
oscillators. The theoretical investigation of biochemical oscillators under the influence of an external periodic signal 
is an appealing direction for future work.   

\appendix

\section{Activator-inhibitor model} \label{app_AI}

The time evolution of $P(n_{R},n_{X},n_{K},n_\mathbf{M},t)$, the probability to find the system in state $(n_{R},n_{X},n_{K},n_\mathbf{M})$ at time $t$, is governed by the chemical master equation
\onecolumngrid
\begin{equation}
\begin{aligned}
\partial_t P(n_{R},n_{X},n_{K},n_\mathbf{M},t) =~&\Big\{k_0[S]n_{M_p}(\epsilon_R^- - 1) + \Omega k_1 [S](\epsilon_R^- - 1)+ k_3 [S] n_{M_p} (\epsilon_X^- - 1) \\
&+ k_2/\Omega\left[n_{X}(n_{R}+ 1)\epsilon_R^+ - n_{X}n_{R}\right] + k_4 [(n_X+1)\epsilon_X^+ - n_X]\\
&+\delta_0/\Omega \left[n_{M_p}(n_R+1)\epsilon_R^+ - n_{M_p}n_R \right] + \delta_1\left[(n_R+1)\epsilon_R^+ - n_R\right]  \\
&+ \delta_2\left[n_{X}[S](\epsilon_R^- - 1)\right]+\delta_3/\Omega \left[n_{M_p}(n_X+1)\epsilon_X^+ - n_{M_p}n_X \right] + \Omega \delta_4 [S] (\epsilon_X^- - 1)\\
&+ a_1/\Omega \left[ (n_{M}+ 1)(n_{R}-n_{MR}+1)\epsilon_M^+\epsilon^-_{MR} - (n_{R}-n_{MR})n_{M}\right] \\
&+ a_2/\Omega \left[ (n_{K}+ 1)(n_{M_p}+ 1)\epsilon_{K}^+\epsilon_{M_p}^+\epsilon^-_{M_pK} - n_{M_p}n_{K}\right] \\
&+ d_1 \left[ (n_{MR}+ 1)\epsilon_M^-\epsilon^+_{MR} - n_{MR}\right] + d_2 \left[ (n_{M_pK}+ 1)\epsilon^-_{K}\epsilon_{M_p}^-\epsilon^+_{M_pK} - n_{M_pK}\right] \\
&+ f_1 \left[ (n_{MR}+ 1)\epsilon_{MR}^+\epsilon^-_{M_p} - n_{MR}\right] + f_2 \left[ (n_{M_pK}+ 1)\epsilon_{K}^-\epsilon^-_{M}\epsilon_{M_pK}^+ - n_{M_pK}\right] \\
&+ f_{-1}/\Omega \left[ (n_{M_p}+ 1)(n_{R}-n_{MR}+1)\epsilon_{MR}^-\epsilon^+_{M_p} - (n_{R}-n_{MR})n_{M_p}\right] \\
&+ f_{-2}/\Omega \left[ (n_{K}+ 1)(n_{M}+ 1)\epsilon_{K}^+\epsilon^+_{M}\epsilon_{M_pK}^- - n_{K}n_{M}\right] \Big\} P(n_{R},n_{X},n_{K},n_\mathbf{M},t),\\
\end{aligned}
\label{eq:AI_CME}
\end{equation}
where we define step operators as $\epsilon_{i}^\pm P(n_{R},n_{X},n_{K},n_\mathbf{M},t)\equiv P(...,n_i\pm1,...,t)$ for $i=R, X , K , M , MR, M_p, M_pK$. 

From the master equation \eqref{eq:AI_CME}, we obtain the equations for the time evolution of the concentrations in the deterministic limit \cite{cao15}, which reads

\begin{equation}
\begin{aligned}
\frac{d[R]}{dt}&=k_0[M_p][S]+k_1[S]-k_2[X][R] - \delta_0[M_p][R] - \delta_1[R] + \delta_2[X][S] \\
\frac{d[X]}{dt}&=k_3[M_p][S] - k_4[X] - \delta_3[M_p][X] + \delta_4[S]\\
\frac{d[M]}{dt}&=f_2[M_pK]+d_1[MR]-a_1[M]([R]-[MR])-f_{-2}[M][K]\\
\frac{d[MR]}{dt}&=a_1[M]([R]-[MR])+f_{-1}[M_p]([R]-[MR])-(f_1+d_1)[MR]\\
\frac{d[M_p]}{dt}&=f_1[MR]+d_2[M_pK]-a_2[M_p][K]-f_{-1}[M_p]([R]-[MR])\\
\frac{d[M_pK]}{dt}&=a_2[M_p][K]+f_{-2}[M][K]-f_2[M_pK]-d_2[M_pK]\\
[K_\text{tot}]&=[K]+[M_pK].\\
\end{aligned}
\label{eqapp:AI_Det}
\end{equation}
Note that $[R]$ is the total concentration of activators and $([R]-[MR])$ is the concentration of free activators which can bind an enzyme M.

\section{KaiC model} \label{app_KaiC}
\onecolumngrid
For the KaiC model, the state of the system is determined by the number of free active KaiC $n_\mathbf{C}=(n_{C_0},n_{C_1},...,n_{C_6})$, inactive KaiC  $n_\mathbf{\widetilde{C}}=(n_{\widetilde{C}_0},n_{\widetilde{C}_1},...,n_{\widetilde{C}_6})$, 
active KaiC bound with a KaiA $n_\mathbf{AC}=(n_{AC_0},n_{AC_1},...,n_{AC_5})$ and free KaiA $n_A$. The time evolution of $P(n_\mathbf{C},n_\mathbf{\widetilde{C}},n_\mathbf{AC},n_A,t)$, the probability to find the system in state $(n_\mathbf{C},n_\mathbf{\widetilde{C}},n_\mathbf{AC},n_A,t)$ at time $t$, is governed by the chemical master equation

\begin{equation}
\begin{aligned}
\partial_t P(n_\mathbf{C},n_\mathbf{\widetilde{C}},n_\mathbf{AC},n_A,t) =~&\Big\{ \sum_{j=0}^5 \gamma \left[(n_{AC_j}+1)\epsilon^+_{AC_j}\epsilon^-_{C_{j+1}} - n_{AC_j} \right] + \gamma K_j^{-1} e^{E/6-\varDelta\mu/2} \left[(n_{C_{j+1}}+1)\epsilon^+_{C_{j+1}}\epsilon^-_{{AC_j}} - n_{C_{j+1}} \right] \\
&+ \sum_{j=0}^5 \widetilde{\gamma} e^{E/6-\varDelta\mu/2} \left[ (n_{\widetilde{C}_j}+1)\epsilon^+_{\widetilde{C}_j}\epsilon^-_{\widetilde{C}_{j+1}} - n_{\widetilde{C}_j} \right] + \widetilde{\gamma} \left[ (n_{\widetilde{C}_{j+1}}+1)\epsilon^+_{\widetilde{C}_{j+1}}\epsilon^-_{\widetilde{C}_j} - n_{\widetilde{C}_{j+1}} \right] \\
&+ \sum_{j=0}^5 \alpha/\Omega \left[(n_{C_j}+1)(n_A+1)\epsilon^+_{C_j}\epsilon^+_{A}\epsilon^-_{AC_j} - n_{C_j}n_A \right] + \alpha K_j \left[(n_{AC_j}+1)\epsilon^+_{AC_j}\epsilon^-_{C_j}\epsilon^-_{A} - n_{AC_j} \right] \\
&+ \sum_{j=0}^6 ke^{\chi E(3-j)/3} \left[(n_{C_j}+1) \epsilon^+_{C_j}\epsilon^-_{\widetilde{C}_j} -  n_{C_j} \right] + ke^{\widetilde\chi E(j-3)/3} \left[(n_{\widetilde{C}_j}+1) \epsilon^+_{\widetilde{C}_j}\epsilon^-_{C_j} -  n_{\widetilde{C}_j} \right] \\
&\Big\}P(n_\mathbf{C},n_\mathbf{\widetilde{C}},n_\mathbf{AC},n_A,t),\\
\end{aligned}
\label{eqapp:Kai_CME}
\end{equation}

where the step operators are defined as as $\epsilon_{j}^\pm P(n_\mathbf{C},n_\mathbf{\widetilde{C}},n_\mathbf{AC},n_A,t)\equiv P(...,n_j\pm1,...,t)$ for $j=C_0, ..., C_6, \widetilde{C}_0, ..., \widetilde{C}_6, AC_0, ...,AC_5,A$.

For the fast binding and unbinding kinetics of KaiA, we introduce coarse-grained active states $n_{C^T_j} = n_{C_j}+n_{AC_j}$, where $j=0,5$ and $n_{C^T_6} = n_{C_6}$ (KaiA does not bind to $C_6$). 
We define the unbound fraction of active states as $p_j=K_j/(K_j+[A])$. The concentration of free KaiA $[A]$ is given implicitly by the constraint 
\begin{equation}
\left[A\right]+\sum_{j=0}^{5} \frac{\left[A\right] \left[C_j^T\right]}{K_{j}+\left[A\right]}  = \left[A_\text{tot}\right].
\label{eqapp:Kai_constraint}
\end{equation}
The state of the coarse-grained system is determined by the total number of active KaiC $n_{\mathbf{C}^T}=(n_{C^T_1},...,n_{C^T_6})$ and inactive KaiC  $n_\mathbf{\widetilde{C}}=(n_{\widetilde{C}_1},...,n_{\widetilde{C}_6})$. The time evolution of $P(n_{\mathbf{C}^T},n_\mathbf{\widetilde{C}},t)$, the probability to find the system in state $(n_{\mathbf{C}^T},n_\mathbf{\widetilde{C}},t)$ at time $t$, is governed by the following chemical master equation
\begin{equation}
\begin{aligned}
\partial_t P(n_\mathbf{C},n_\mathbf{\widetilde{C}},t) =~&\Big\{ \sum_{j=0}^5 (1-p_j)\gamma \left[ (n_{C^T_j}+1)\epsilon^+_{n_{C^T_j}}\epsilon^-_{n_{C^T_{j+1}}} -n_{C^T_j}  \right] \\
&+ \sum_{j=0}^4 [A]p_j\gamma K_j^{-1}e^{E/6-\varDelta\mu/2} \left[(n_{C^T_{j+1}}+1)\epsilon^+_{n_{C^T_{j+1}}}\epsilon^-_{n_{C^T_j}} -n_{C^T_{j+1}}  \right] \\
&+ \gamma K_5^{-1}e^{E/6-\varDelta\mu/2} \left[(n_{C^T_{6}}+1)\epsilon^+_{n_{C^T_{6}}}\epsilon^-_{n_{C^T_5}} -n_{C^T_{6}}  \right] \\
&+ \sum_{j=0}^5 \widetilde{\gamma} e^{E/6-\varDelta\mu/2} \left[ (n_{\widetilde{C}_j}+1)\epsilon^+_{\widetilde{C}_j}\epsilon^-_{\widetilde{C}_{j+1}} - n_{\widetilde{C}_j} \right] + \widetilde{\gamma} \left[ (n_{\widetilde{C}_{j+1}}+1)\epsilon^+_{\widetilde{C}_{j+1}}\epsilon^-_{\widetilde{C}_j} - n_{\widetilde{C}_{j+1}} \right] \\
&+ \sum_{j=0}^6 p_j ke^{\chi E(3-j)/3} \left[(n_{C_j}+1) \epsilon^+_{C_j}\epsilon^-_{\widetilde{C}_j} -  n_{C_j} \right] + ke^{\widetilde\chi E(j-3)/3} \left[(n_{\widetilde{C}_j}+1) \epsilon^+_{\widetilde{C}_j}\epsilon^-_{C_j} -  n_{\widetilde{C}_j} \right] \\
&\Big\}P(n_\mathbf{C},n_\mathbf{\widetilde{C}},t).\\
\end{aligned}
\label{eqapp:Kai_CGCME}
\end{equation}

From the coarse-grained master equation \eqref{eqapp:Kai_CGCME}, we obtain the equations for the time evolution of the concentrations in the deterministic limit
\begin{equation}
\begin{aligned}
\frac{d\left[C_j^T\right]}{dt} =&~(1-p_{j-1})\gamma \left[C_{j-1}^T\right] + p_{j+1}\gamma K_{j+1}^{-1} \e^{E/6-\varDelta\mu/2}\left[A\right]\left[C_{j+1}^T\right] -(1-p_j)\gamma \left[C_{j}^T\right](1-\delta_{j,6}) \\
& - p_j\gamma K_j^{-1} \e^{E/6-\varDelta\mu/2}\left[A\right] \left[C_{j}^T\right](1-\delta_{j,0})  + k\e^{\widetilde{\chi}E(3-j)/3}\left[\widetilde{C}_j\right] - p_jk\e^{\chi E(j-3)/3}\left[C_j^T\right] \\
\frac{d\left[\widetilde{C}_{j}\right]}{dt} =&~\widetilde{\gamma} \left( \left[\widetilde{C}_{j+1}\right] - \left[\widetilde{C}_j\right] (1-\delta_{j,0}) \right)  - \widetilde{\gamma} \e^{\varDelta\mu/2} \left( \left[\widetilde{C}_{j-1}\right] - \left[\widetilde{C}_j\right](1-\delta_{j,6}) \right) \\	
& + p_jk\e^{\chi E(j-3)/3}\left[C_j\right] - k\e^{\widetilde{\chi}E(3-j)/3}\left[\widetilde{C}_j\right], \\
\end{aligned}
	\label{eqapp:Kai_det}
\end{equation}
where $[A]$ is given by Eq.~\eqref{eqapp:Kai_constraint}.
\vspace{1cm}
\twocolumngrid

%


\end{document}